\def\l{$\ell$}
\def\m{$m$}
\def\n{$n$}

\def\ahat{{\mathaccent "7E {\cal A}}}
\def\bhat{{\mathaccent "7E {\cal B}}}

\def\dhat{{\mathaccent "7E {\cal D}}}
\def\shat{{\mathaccent "7E {\cal S}}}
\def\chat{{\mathaccent "7E {\cal C}}}

\def\ket{\vert \vert  \{ \emptyset \} \rangle}
\def\ket2{\vert \vert \otimes \{ R \} \rangle}

\def\no{\noindent }
\def\mbf#1{{\mathbf {#1}}}

\def\eq{\ =\ }
\def\mns{\ -\ }
\def\pls{\ +\ }
\def\be{\begin{equation}}
\def\ee{\end{equation}}
\def\<{[}
\def\>{]}
\def\citesep{, }

\documentclass[12pt,epsfig,a4paper]{iopart}
\usepackage{epsfig}
\begin{document}
\title[Symmetry reduction in the TB-LMTO-ASR  for random binary alloys]
{Symmetry reduction in the  augmented space recursion formalism for random binary alloys}
\author{\bf Kamal Krishna Saha \footnote{Email: kamal@bose.res.in}, Tanusri Saha-Dasgupta and Abhijit Mookerjee}
\address { S. N. Bose National Centre for Basic Sciences.
Block-JD, Sector-III, Kolkata 700098, India.}
\author{Indra Dasgupta}
\address{Department of Physics, Indian Institute of Technology, Powai, Mumbai 400076, India}

\begin{abstract}{We present here an efficient method which systematically reduces the rank
of the augmented space and thereby helps to implement augmented space recursion for any real 
calculation. Our method is based on the symmetry of the Hamiltonian in the augmented space
and keeping recursion basis vectors in the irreducible subspace of the Hilbert space.}
\end{abstract}
\parindent 0pt

\section{Introduction}

In a series of earlier communications \cite{asr,asr2} we have proposed the augmented space recursion (ASR) 
as an efficient computational technique for the calculation of configuration averaged electronic properties
of random binary alloys. The method is a combination of the augmented space formalism \cite{asf}
and the recursion method of Haydock \etal \cite{hhk}. When coupled with the local density functional approximation based tight-binding
linearized muffin-tin orbital \cite{lmto1,lmto2} it provides a first-principles determination of electronic
structure and total energy calculations for disordered alloys. Effects such as short-ranged ordering and  local lattice distortions
due to size mismatch of the constituent atoms can very easily be incorporated in the methodology \cite{sro1,sro2,
latdis}. Recently the method has been applied to study dispersion and line widths for phonons in disordered alloys.
Since recursion can take into account disorder in beyond nearest-neighbour force constants as well as large
environmental effects, this application was a step forward in the theory of phonons in disordered alloys \cite{phonons}.
It has been pointed out repeatedly in the above publications that the ASR with its terminator approximation \cite{bp,ln}
goes beyond the standard single-site mean field theories and accurately takes off-diagonal disorder into account.
A number of attempts have been made for practical implementation of mean-field theories beyond the single site
approximation. One of the
most sophisticated developments among the mean-field theories, is the travelling cluster approximation 
of Gray and Kaplan \cite{KG1,KG2} and its latest version  : the itinerant coherent potential approximation (ICPA) \cite{subhra},
where a small number of classes of selected excitations in augmented space are preserved. In
spite of the  simplification so introduced, the resulting equations are still  
formidable enough to discourage further generalization to larger clusters beyond the nearest
neighbour. In the cluster CPA proposed by 
Mookerjee \cite{ccpa1,ccpa2}, in which one retains disorder fluctuations only from a chosen cluster
and replaces the rest of the system by an effective medium,
one is restricted by the limitation of solving number of coupled self-consistent equations. The number of 
equations increases with the size of chosen cluster, which again discouraged extension to larger clusters
beyond the nearest neighbour. This brings us back to the attempt at the implementation of the ASR for
large environment fluctuations, implying accurate evaluation of a large sequence of recursion coefficients. We shall discuss this in some detail in the following section.

The ASR, in its practical implementation to real alloy systems, has several practical problems. 
The formalism
maps a disordered Hamiltonian, described in a Hilbert space spanned by a countable basis in which
preferably its representation is sparse, onto an ordered Hamiltonian in an 
enlarged space. This enlarged or {\sl augmented} space 
is constructed by taking the outer product of  the configuration space
of the random variables of the disordered Hamiltonian together with original Hilbert space
in which the disordered Hamiltonian was described. 
The main difficulty in the implementation of recursion in this augmented space is its enormous rank. 
A system of $N$ sites with binary distribution has an 
augmented  space of rank $N\times 2^N$. 
This enormous rank of the augmented space and its description on the computer is what had discouraged
earlier workers in implementing the ASR, although the augmented space formalism was first proposed about
thirty years ago \cite{asf}.  The second problem was the recursive propagation of errors. Very
small errors during recursion (inevitable in numerical implementations) tends to add up leading to loss
of orthogonality in the recursion basis. This leads to appearance of ghost states. Haydock and Te \cite{ht} has
overcome this difficulty in proposing the {\sl dynamical recursion}. We refer the reader to their paper for
details.

In this communication we shall describe the reduction of the effective rank of the augmented space on which the
recursion is carried out, using the local point-group symmetries in both the Hilbert space and the configuration space. Such reduction is well known in various manipulations in reciprocal space for ordered systems. However,
such a reduction has not been implemented in real space recursion, although it had been proposed by
Gallagher \cite{Ga} quite some time ago.
It had been shown by Gallagher that if we start recursion
with a state belonging to an irreducible subspace of a Hilbert space, subsequent recursion 
always stays within this subspace. This allows us significant reduction in rank of the required
subspace and makes this method practically feasible. Further, this method with its terminator
approximation, retains the Herglotz
properties of the configuration averaged Green function.

\section{The augmented space recursion}

\subsection{The recursion method}

The recursion method addresses inversions of infinite matrices 
\cite{hhk}. Once a sparse representation of an operator in Hilbert space, 
$\tilde{H}$, is known in a countable basis,
the recursion method obtains an alternative basis in which the operator
becomes tri-diagonal. This basis and the representations of the operator in it are found recursively
starting from $\vert  u_1\rangle = \vert R\rangle \otimes \vert {\mathrm 1}\rangle$ :

\begin{eqnarray}
\vert u_2\rangle & = & \tilde{H}\vert u_1\rangle - \alpha_1 \vert  u_1\rangle \nonumber\\
\vert u_{n+1}\rangle & = & \tilde{H}\vert u_n\rangle - \alpha_n \vert u_n \rangle - \beta^2_n \vert u_{n-1}\rangle \quad\mbox{n $\geq$ 2}  
\end{eqnarray}

\no where,

\begin{equation}
\alpha_n \ = \ \frac{\langle u_n\vert {\tilde H}\vert u_n\rangle}{\langle u_n\vert u_n\rangle} \quad\quad
\beta_n^{2} \ = \ \frac{\langle u_{n}\vert u_{n}\rangle}{\langle u_{n-1}\vert u_{n-1}\rangle}  \end{equation}

\no for n$\geq$ 2. In this basis the representation of $\tilde{H}$ is tridiagonal : 

\begin{eqnarray*}
\fl \frac{\langle u_n\vert \tilde{H}\vert u_n\rangle}{\langle u_n\vert u_n\rangle} \ =\ \alpha_n \quad ; \quad
\frac{\langle u_{n-1}\vert  \tilde{H}\vert u_{n}\rangle}{\left[\langle u_{n-1}\vert u_{n-1}\rangle\langle
 u_{n}\vert u_{n}\rangle\right]^{1/2}}\ =\ \beta_n \enskip ;\enskip
\langle u_n\vert \tilde{H}\vert u_{m}\rangle \ =\  0 \quad m\geq n+2
\end{eqnarray*}

\no and the averaged Green function can be written as a continued fraction :

\begin{eqnarray}
\ll G_{RR}(E)\gg  & = &\langle u_1\vert (E\tilde{I}-\tilde{H})^{-1}\vert  u_1\rangle \nonumber\\
                  & & \phantom{x}\nonumber\\
&= & \frac{1}{\displaystyle{E-\alpha_1-\frac{\beta_2^2}{\displaystyle{E-\alpha_2-\frac{\beta_3^2}
{\displaystyle{\ddots\atop{\displaystyle \quad\quad E-\alpha_N-T(E)}}}}}}}\nonumber\\
\end{eqnarray}

The asymptotic part of the continued fraction is obtained from the initial set of coefficients
$\{\alpha_n,\beta_n\}$ for $n < N$, using the idea of {\sl terminators} \cite{bp,ln}. For small values
of $N$ we have large inaccuracies and the more the structure in the spectral density of $\tilde{H}$
the larger is the $N$  needed to maintain the window of accuracy required by us.

\subsection{The TB-LMTO Hamiltonian for random binary alloys}

In earlier communications we have described how to deal with random binary alloys \cite{asr}
within the framework of the tight-binding linearized muffin-tin orbital method (TB-LMTO)
\cite{lmto1,lmto2} and using the augmented space formalism (ASF). We refer the readers to the seminal papers mentioned earlier for details. Here we shall introduce only the salient
features which will be required by us subsequently in this communication. We begin by 
setting up a muffin-tin potential with centres at the atomic sites $R$ on a lattice.
Next, we inflate the muffin-tins into atomic spheres and start from a most-localized TB-LMTO
representation of the Hamiltonian~:

\[ {\mathbf H}  =  \sum_{RL} {\bf C}_{RL} \enskip {\cal P}_{RL} +
\sum_{RL}\sum_{R'L'}\Delta_{RL}^{1/2} S_{RL,R'L'} 
\Delta_{R'L'}^{1/2} \enskip {\cal T}_{RL,R'L'} \]

The {\sl potential parameters}, ${\bf C}$ and $\Delta$, are diagonal
matrices in the angular momentum indeces, and

\begin{eqnarray}
{\bf C}_{RL}& = &  C_L^A\  n_{R} + C_L^B\ \left( 1-n_{R} \right) \nonumber \\
\Delta_{RL}^{1/2} & = & \left( \Delta_L^A \right)^{1/2}
n_{R} + \left( \Delta_L^B \right)^{1/2}\left( 1-n_{R} \right) 
\end{eqnarray}

Here $L$ is a composite label $\{\ell,m,m_s\}$ for the angular momentum quantum numbers.
$n_R$ is the random site-occupation
variable which takes values 0 or 1 with probability $x$ or $y$ $(x+y=1)$ respectively, 
depending upon whether the muffin-tin labelled by $R$ is occupied by $A$ or 
$B$-type of atom.

The ${\cal P}_{RL}$ and ${\cal T}_{RL,R'L'}$ are projection and transfer 
operators in the Hilbert space ${\cal H}$ spanned by the tight-binding basis 
$\left\{\vert RL \rangle \right\}$.

\subsection{The augmented space formalism within the TB-LMTO}

We can associate the random variable $n_R$ with an operator ${\cal M}_R$ whose eigenvalues
correspond to the observed values of $n_R$, and whose corresponding eigenvectors, 
$\{\vert 0_R\rangle, \vert 1_R \rangle \}$ are the {\sl state vectors} of the variable. 
These state vectors of the set of $N$ random variables $n_R$ of rank $2^N$ span a space called
configuration space $\Phi$ with configurations of the kind $\vert \uparrow\uparrow\downarrow\ldots\downarrow\uparrow\ldots\rangle$ where 

\begin{eqnarray*}
\vert\uparrow_R\rangle &\eq& \sqrt{x}\ \vert {0_R} \rangle \pls \sqrt{y}\ \vert {1_R} \rangle \\
\vert\downarrow_R\rangle &\eq& \sqrt{y} \ \vert {0_R} \rangle \mns \sqrt{x}\ \vert {1_R} \rangle 
\end{eqnarray*}

If we define the configuration $\vert\uparrow\uparrow\ldots\uparrow\ldots\rangle$ as the  {\sl reference} configuration,
then any other configuration may be uniquely labelled by the {\sl cardinality sequence} : 
$\{R_k\}$,  which is the sequence of
positions where we have a $\downarrow$ configuration. The {\sl cardinality sequence} of the {\sl reference} configuration
is the null sequence $\{\emptyset\}$

The augmented space theorem states that

\be \ll A(\{n_R\}) \gg \eq < \{\emptyset\}\vert \widetilde{{\mbf A}}\vert \{\emptyset\}> \ee

\no where,

\[ \widetilde{\mbf A} \eq \int \ldots \int A(\{\lambda_R\})\ \prod d{\mbf P}(\lambda_R) \quad \epsilon\quad \Phi \]

\no {\bf P}($\lambda_R$) is the spectral density of the self-adjoint operator ${\cal M}_R$, which is
such that the probability density of $n_R$ is given by : 

\[ p(n_R)\eq -\frac{1}{\pi}\ \lim_{\delta\rightarrow 0}\ \Im m\ \langle \uparrow _R\vert \left(\rule{0mm}{4mm}(n_R+i\delta)
{{\mbf I}}\mns {{\cal M}}_R\right)^{-1}\vert\uparrow_R\rangle \]

${\cal M}_R$ is an operator in the space of configurations $\psi_R$ of the variable 
$n_R$. This is of rank 2 and is spanned by the {\sl states} $ \{ \vert \uparrow_R \rangle,
\vert \downarrow_R \rangle \} $.

\begin{equation}
{{\cal M}}_R \enskip =\enskip x \enskip {\cal P}_{\uparrow}^R \enskip +\enskip
  y\enskip {\cal P}_{\downarrow}^R\enskip +\enskip \sqrt{xy}\ \left( {\cal
  T}_{\uparrow \downarrow}^R + {\cal T}_{\downarrow\uparrow}^R \right)
  \end{equation}

The expended Hamiltonian ${\widetilde{\bf H}}$ in the augmented space is constructed by 
replacing all the random variables $n_R$ by the corresponding operators 
${\cal M}_R$. It is an operator in the augmented space 
$\Psi$ = ${\cal H} \otimes \Phi$. So the ASF maps 
a disordered Hamiltonian described in a Hilbert space ${\cal H}$ onto an ordered Hamiltonian 
in an enlarged space $\Psi$, where the space $\Psi$ is constructed by augmenting the 
configuration space $\Phi=\prod_R^{\otimes}\psi_R$ of the random variables of 
the disordered Hamiltonian together with the Hilbert space ${\cal H}$ of the disordered
Hamiltonian. The  configuration space $\Phi$ is of rank 2$^{N}$ if there are N muffin-tin 
spheres in the system. 

We may rewrite the expression for the configuration average of the Green operator as :

\begin{equation}
\ll {\mathbf G(E)}\gg \eq  <\{\emptyset\}\vert \dhat\ 
\left( \ahat+\bhat+\chat-\shat\right)^{-1}
\dhat \vert  \{ \emptyset \} >
\label{diag3}
\end{equation}

where, 
\begin{eqnarray*}
\ahat & = & \sum_{RL}  A\left[\rule{0mm}{4mm}(E-C_L)/\Delta_L\right]\enskip {\cal P}_R 
\otimes {\cal P}_{L} \otimes {\cal I} \\
\bhat & = & \sum_{RL} B\left[\rule{0mm}{4mm} (E-C_L)/\Delta_L \right] \enskip 
   {\cal P}_{R} \otimes {\cal P}_{L} \otimes {\cal P}_{\downarrow}^R \\
\chat & = & \sum_{RL} C\left[\rule{0mm}{4mm} (E-C_L)/\Delta_L \right] \enskip {\cal P}_{R} 
\otimes {\cal P}_{L}\otimes \left\{ {\cal T}^R_{\uparrow\downarrow} + {\cal
  T}^R_{\downarrow\uparrow} \right\}  \\
\shat & = &  \sum_{RL} \sum_{R'L'} S_{RL,R'L'} \enskip {\cal T}_{RR'} 
\otimes {\cal T}_{LL'} \otimes  {\cal I}  \\
 \phantom{x}& & \\
\dhat & = & \sum_{RL} A\left( \Delta^{1/2}_L \right)\enskip {\cal P}_R 
\otimes {\cal P}_{L} \otimes {\cal I} + \sum_{RL} B\left( \Delta^{1/2}_L \right)
  \enskip {\cal P}_{R}\otimes {\cal P}_{L}\otimes {\cal P}_{\downarrow}^R \ldots \\
   & &  +
  \sum_{RL} C\left( \Delta^{1/2}_L \right)
  \quad {\cal P}_{R} \otimes {\cal P}_{L}\otimes \left\{ {\cal T}^R_{\uparrow\downarrow} + {\cal
  T}^R_{\downarrow\uparrow} \right\} 
\end{eqnarray*}

\no and

\begin{eqnarray*}
  A(V) & = &  x V_A + y V_B \quad i.e. \mbox{ the average of }V \\
  B(V) & = &  (1-2x) (V_A - V_B) \\
  C(V) & = & \sqrt{xy} (V_A - V_B)
  \end{eqnarray*}

Since,

\[ \dhat \vert  \{ \emptyset \} \rangle \enskip =\enskip
  A\left(\Delta_L^{-1/2} \right) \vert \{ \emptyset \}
  \rangle + C\left( \Delta_L^{-1/2} \right)  \vert \{R\}
  \rangle \enskip =\enskip \vert \mbox{\rm 1} \} \nonumber
\]

  The ket $\vert 1 \} $ is not normalized. We first write the above
  in terms of a normalized ket $\vert 1 \rangle $ =
  $\left[A(1/\Delta)\right]^{-1/2} \vert  1 \} $. We  now have :

\begin{eqnarray}
\ll G(E) \gg & \eq & \langle 1 \vert \left(E - \ahat^{\prime} - \bhat^{\prime} -
  \chat^{\prime} -
  \shat^{\prime} \right)^{-1} \vert 1 \rangle \nonumber\\
& \eq & \langle 1 \vert \left( E\mns {\tilde{H}}^{eff}\right)^{-1}\vert 1\rangle \nonumber\\
{\phantom{xxxx}\tilde{H}}^{eff} & \eq & \ahat^\prime +\bhat^\prime +\chat^\prime + \shat^\prime 
\label{kasr}
\end{eqnarray}

where,

\begin{eqnarray}
\fl  \ahat^{\prime}  =  \sum_{RL}\left\{ A\left( C_L/\Delta_L \right) / A\left(1/\Delta_L 
\right)\right\} \enskip {\cal P}_R \otimes {\cal P}_L \otimes\tilde{\cal I} \nonumber\\
\fl \bhat^{\prime}  =  \sum_{RL}\left\{ B\left[ (C_L-E)/\Delta_L\right] / 
A\left( 1/\Delta_{L}
  \right)\right\}\enskip  
   {\cal P}_{R} \otimes {\cal P}_L\otimes {\cal P}_{\downarrow}^{R} \nonumber\\
\fl \chat^{\prime}  = \sum_{RL} \left\{C\left[ (C_L-E)/\Delta_L\right] / A\left( 1/\Delta_L
  \right)\right\}\enskip  
  {\cal P}_{R} \otimes {\cal P}_{L} \otimes \left\{ {\cal T}^{R}_{\uparrow\downarrow} + {\cal
  T}^{R}_{\downarrow\uparrow} \right\}\nonumber  \\
\fl  \shat^{\prime}  =  \sum_{RL} \sum_{R'L'}\left[A\left( 1/\Delta_L
\right)\right]^{-1/2} \ S_{RL,R'L'} \
\left[ A\left( 1/\Delta_{L'} \right)\right]^{-1/2} {\cal T}_{RR'}\otimes{\cal T}_{LL'} 
\otimes {\mathaccent "7E {\cal I}}\nonumber\\
\label{kasr2}
  \end{eqnarray}

This equation is now exactly in the form in which the recursion method may now be applied. 
The computational burden is considerably reduced due to this diagonal formulation,
the recursion now becomes energy dependent as is clear from the form of the 
 effective Hamiltonian as shown in \eref{diag3} and discussed in \cite{diag1,diag2}. The recursion formalism of the ordered Hamiltonian was
free of this constraint. This energy dependence makes the recursion technique 
computationally unsuitable because to obtain the Green functions we have to carry out
recursion per energy point of interest. This problem has been tackled using {\sl seed
recursion technique} \cite{diag2}. The idea is to choose a few seed points across the
energy spectrum uniformly, carry out recursion over those points and then interpolate
the values of coefficients across the band. In this way one may reduce computation
time. For example, if one is interested in an energy spectrum of 200 points, in the
bare diagonal formulation recursion has to be carried out at all the 200 points but
in the seed recursion technique one needs to perform recursions only at 10-15 points.
The whole idea stems from the fact that in most of the cases of binary alloys, it is
seen that the recursion coefficients $a_n$ and $b_n$ vary quite weakly
across the energy spectrum.
At this point we note that the above expression for the averaged $\ll G_{LL}(E)\gg$ is 
{\sl exact}. 

\section{Symmetry reduction of the augmented space rank}

We mentioned earlier that recursion on the augmented space is not computationally feasible because
of its large rank. For a binary alloy with $N$ sites and with
only $s$-orbitals on them, the rank of the augmented space is $N\times 2^N$. 
Implementing recursion on this huge space for a sufficient number of steps to ensure accuracy
is often not feasible on available computers.  However, if we exploit
the symmetry both of the underlying lattice in real space and of the configuration space 
(which arises due to the homogeneity of the disorder and arrangements of atoms on the underlying
lattice), 
the rank of the irreducible part of the augmented space in which the recursion is effectively carried out becomes
tractable. The conceptual advantages in ASF include apart from analyticity, translational
and underlying lattice symmetries automatically built into the augmented space Hamiltonian.
This allows us to involve the idea of utilizing symmetry operations present in 
both the real and configuration spaces, in the context of recursion
method, reducing the rank of Hamiltonian drastically and making the implementation
of ASF feasible.  Since the augmented space recursion
essentially retains all the properties of real space recursion but is described in a
much enlarged space, it will be useful to consider symmetry operations in real space
recursion first and then to consider those in the full augmented space. 

During the recursion, the basis member $|u_n\rangle$ is generated from the starting
state $|u_1\rangle$ by repeated application of the Hamiltonian. If the starting state
belongs to an irreducible sub-space of the Hilbert space then all subsequent states
generated from recursion will belong to the same irreducible sub-space. Physically, we
may understand this as follows: the recursion states $|u_n\rangle$ carry information 
of distant environment of the starting state. If the Hamiltonian is nearest-neighbour
{\sl only}, then the state $|u_2\rangle$, which arises by the application of the Hamiltonian on $\vert u_1\rangle$
is a combination of states in the nearest shell with which $|u_1\rangle$ couple via
the Hamiltonian. Similarly, $|u_n\rangle$ is a combination of n-th neighbour shell with
which $|u_1\rangle$ is coupled via the Hamiltonian. If $\cal R$ is a point group symmetry
of the Hamiltonian, then all $n$-th neighbour-shell states which are related to one another
through the symmetry operator must have equal coupling to $|u_1\rangle$. Hence, it is
useful to consider among the $n$-th neighbour-shell states of which $|u_n\rangle$ is a
linear combination, only those belonging to the irreducible subspace and redefine the
Hamiltonian operation.

\begin{figure}[t]
\centering
\epsfxsize=3.0in \epsfysize=3in \rotatebox{0}{\epsfbox{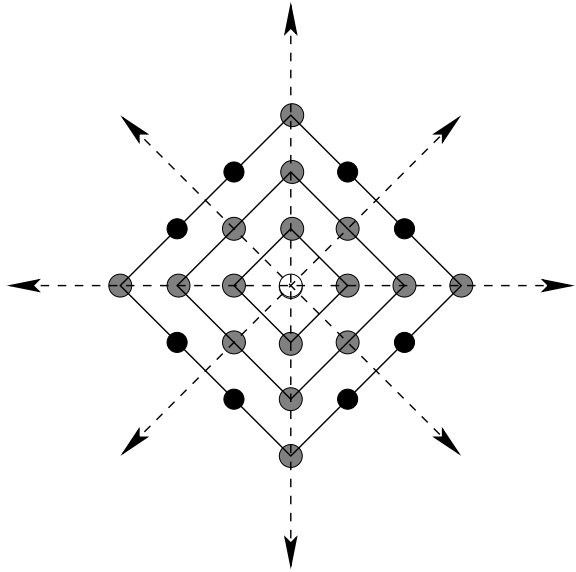}}
\caption{Nearest neighbour shells around a central site on a square lattice.
The colour coding indicates the symmetry weight of the sites : white represents
sites with weight 1, gray with weight 4 and black with weights 8.}
\label{fig1}
\end{figure}

As an example, take a nearest-neighbour $s$-state Anderson model on a square lattice,
 with a binary
distribution of its diagonal elements. We shall label the tight-binding 
basis by the position of the lattice points in units of the lattice constant : e.g.
$|(mn)\rangle$ where $m,n$ are integers $0,\pm 1,\pm 2,\ldots $. The starting state 
$|(00)\rangle$ belongs to the one dimensional representation of the
point group of a square lattice. This state then couples with linear combinations of states
on neighbour shells which are symmetric under square rotations~:

\begin{eqnarray}
|(0m)\} &=& \left(\rule{0mm}{4mm} |(0m)\rangle+|(m0)\rangle+|(\bar{m}0)\rangle+|(0\bar{m})\rangle \right)/2,\ \ m>0 \nonumber\\
|(mm)\} &=& \left(\rule{0mm}{4mm} |(mm)\rangle+|(\bar{m} m)\rangle+|(m\bar{m})\rangle+|(\bar m\bar{m})\rangle \right) /2,
 \ \ m>0\nonumber\\
|(mn)\} &=& \left(\rule{0mm}{4mm}|(mn)\rangle+|(\bar{m}n)\rangle+|(m\bar{n})\rangle+|(\bar{m}\bar{n})\rangle\right. \ldots \nonumber\\
&&\!\!\!\!\!\!+ \left.|(nm)\rangle+|(\bar{n}m)\rangle+|(n\bar{m})\rangle+|(\bar{n}\bar{m})\rangle\rule{0mm}{4mm}\right)/(2\sqrt 2)\nonumber\\
& &  \quad\quad\quad\quad\quad \quad\quad\quad\quad \quad\quad\quad\quad \quad\quad  m>1,\ 0<n<m  
\label{eqs}
\end{eqnarray} 

The \fref{fig1} shows the grouping together of sites with the local symmetry of the square lattice on the first three nearest neighbour shells of 
 the central site. The first and second groups in \eref{eqs} are coloured gray in
\fref{fig1} and the last group coloured black. The labels on the groups (shown on the left sides of the
equations \eref{eqs}) are confined {\sl only} to the upper right quadrant of the lattice. 

If we go up to $N$ shells (for large $N$) there are about $2N^2$ states in the diamond 
shaped nearest-neighbour cluster. However, there are only $(N^2/4+N/2)\sim N^2/4$ states
with square symmetry.

So within this reduction we can work only in 1/8-th  of the lattice, provided  we
incorporate proper weights to the states to reproduce the correct matrix elements.  
If $|R\rangle$ and $|R'\rangle$ are two states
coupled to each other via the Hamiltonian, and both belong to the same irreducible subspace,
and if $|R_1\rangle,|R_2\rangle,\ldots,|R_{\cal{W}_R}\rangle$ are states obtained by operating
on $|R\rangle$ by the symmetry group operations of the real space lattice, then ${\cal W}_R$ is called
the {\sl weight} associated with the state labelled by $R$. If we wish to retain only the
states in the irreducible subspace and throw out the others and yet obtain the same results,
we have to redefine the Hamiltonian matrix elements as follows~:

\begin{equation}
\langle R|H|R'\rangle \ \rightarrow \ \sqrt{\frac{{\cal W}_{R'}}{{\cal W}_R}} \ \langle R|H|R'\rangle
\end{equation}

\begin{figure}
\centering
\epsfxsize=5.4in \epsfysize=4in \epsfbox{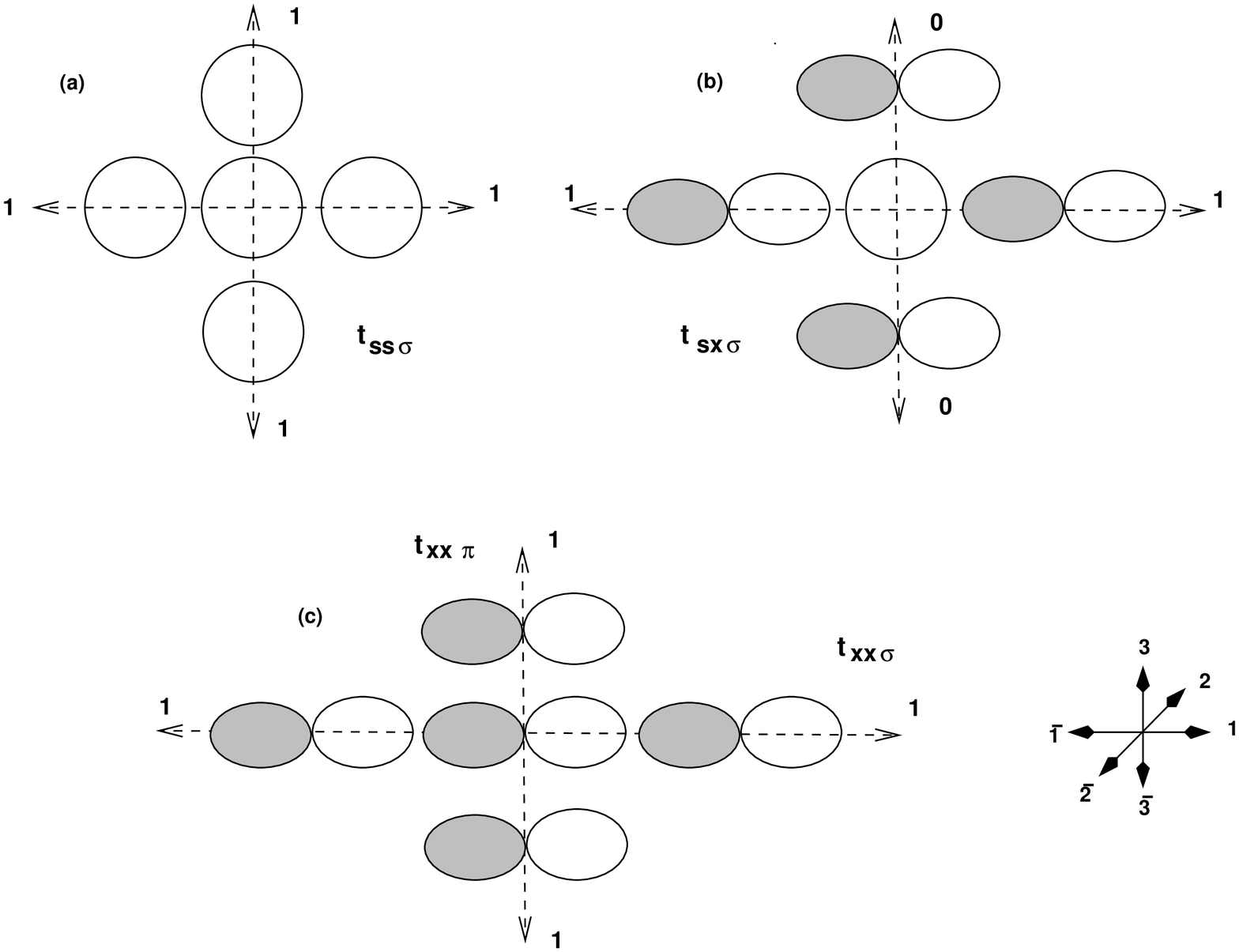}
\caption{$sp$-orbitals on a cubic lattice. Only projections on the $xy$ plane are shown. The positive
parts of the lobes are unshaded, while the negative parts of the lobes are shaded. The values of the
symmetry factor for the overlaps ($\beta_{LL'}$(\l\m\n), defined in the text) are shown for each direction.
The numbering of sites on the cubic lattice is shown in the inset of the right bottom.}
\label{fig2}
\end{figure}

\begin{figure}[h]
\centering
\epsfxsize=6.5in \epsfysize=6.in \rotatebox{270}{\epsfbox{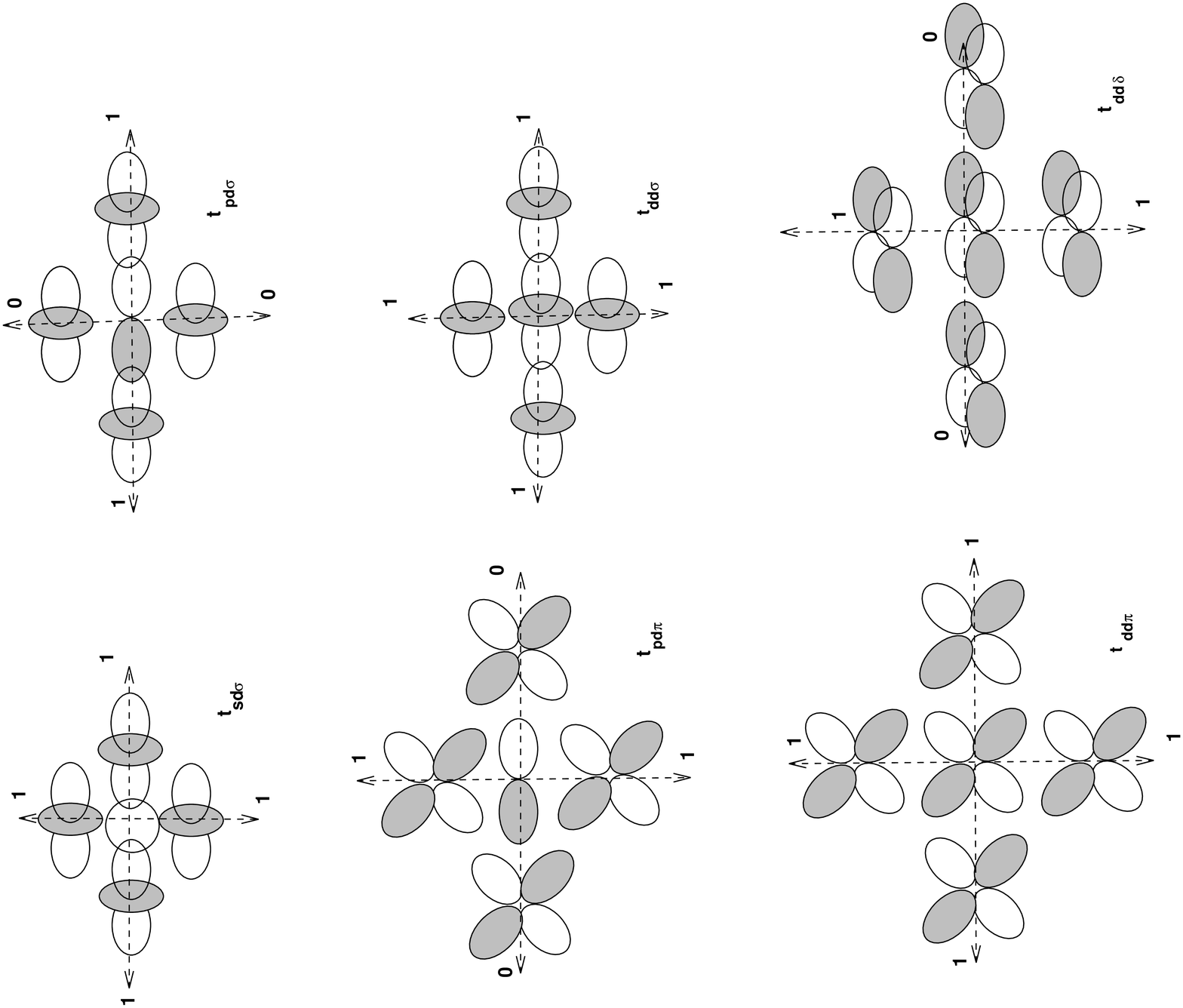}}
\caption{Overlaps involving $d$ orbitals in the $e_g$ and $t_{2g}$ symmetries.  Only projections on the $xy$ plane are shown. The positive
parts of the lobes are unshaded, while the negative parts of the lobes are shaded.}
\label{fig3}
\end{figure}

In the above prescription, the new irreducible basis introduced by us, reflects only the symmetry
of the underlying lattice and holds good for a model system which has $s$-like orbitals only.
But for a real system, the TB-LMTO minimal basis contains members with other  symmetries
as well. For example, in a cubic lattice with a $spd$ minimal basis, we have basis members with
$s$, $p$, $e_g$ and $t_{2g}$ symmetries.   The symmetry of the
orbitals is reflected in the two-centered Slater-Koster integrals and this prohibits overlap
integrals at certain positions, called {\sl symmetric positions} with respect to the overlapping
orbitals. A few of these symmetric positions for a simple cubic lattice are shown in figure \ref{fig2}. For the sake of clarity we have shown only the projections on $x-y$ plane. 
We have indicated the positive and negative parts of the orbital lobes by different shades. 
From \fref{fig2}(b) it is easy to argue that since the Hamiltonian is spherically symmetric and the product 
$\phi_s(000)\phi_{p_x}(010)$ is positive in the upper right quadrant and negative in the
upper left quadrant : $\langle 0,s\vert H \vert 2,p_x\rangle \ =\ 0$. The same is
true for  $\langle 0,s\vert H \vert \bar{2},p_y\rangle \ =\ 0$. Similarly, 
 $\langle 0,s\vert H \vert 3,p_z\rangle \ =\ \langle 0,s\vert H \vert \bar{3},p_z\rangle \ =\ 0$. On the other hand,
 $\langle 0,s\vert H \vert 1,p_x\rangle \ =\ -  \langle 0,s\vert H \vert \bar{1},p_x\rangle \neq 0$.

\begin{table}[p]
\centering
\begin{tabular}{||l|l|l|l||}\br
Matrix  & Slater-Koster  & Symmetry  & Condition for it \\ 
Element               & Expression               & Factor                 & to be zero         \\ \br
$t_{ss}$  & $t_{ss\sigma}$ & $\beta^{\ell mn}_{s,s}$ & -- \\
$t_{s,x}$  & $\rule{0mm}{4mm}\ell$\ $t_{sp\sigma}$ & $\beta^{\ell mn}_{s,p_x}$ & \l = 0 \\
$t_{s,xy}$   & \rule{0mm}{5mm}$\sqrt{3}\ $\l\m\ $t_{sd\sigma}$ & $\beta^{\ell mn}_{s,xy}$ & \l =0 or \m=0 \\
$t_{s,x^2-y^2}$   & $\rule{0mm}{5mm}\sqrt{3}/2\ $(\l$^2$-\m$^2$)\ $t_{sd\sigma}$ & $\beta^{\ell mn}_{s,x^2-y^2}$ &  \l=\m \\
$t_{s,3z^2-r^2}$   & \rule{0mm}{5mm}[$n^2$-1/2(\l$^2$+\m$^2$)]\ $t_{sd\sigma}$ & $\beta^{\ell mn}_{s,3z^2-r^2}$ & $n=1/\sqrt{3}$ \\ \mr
$t_{x,x}$   & $\rule{0mm}{5mm}$\l$^2$\ $t_{pp\sigma}$+(1-\l$^2$)\ $t_{pp\pi}$ & $\beta^{\ell mn}_{x,x}$ & --  \\
$t_{x,y}$   & $\rule{0mm}{5mm}$\l\m\ ($t_{pp\sigma} - t_{pp\pi}$) & $\beta^{\ell mn}_{x,y}$ & \l=0 or \m = 0 \\
$t_{x,xy}$   & $\rule{0mm}{5mm}$\ \m\ [$\sqrt{3}$\l$^2$\ $t_{pd\sigma}$ + (1-2\l$^2$) $t_{pd\pi}$] & $\beta^{\ell mn}_{x,xy}$ & \m = 0 \\
$t_{x,yz}$   & $\rule{0mm}{5mm}\sqrt{3}$\ \l\m\n\ ($t_{pd\sigma}$ -2\ $t_{pd\pi}$) & $\beta^{\ell mn}_{x,yz}$ &\l=0 or \m=0 $\cdots$ \\
  &  &  & $\cdots$  or \n = 0 \\
$t_{x,x^2-y^2}$   & \rule{0mm}{5mm}\ \l\ [$\sqrt{3}/2$\ (\l$^2$-\m$^2$)\ $t_{pd\sigma}$ +(1-\l$^2$+\m$^2$) $t_{pd\pi}$] & $\beta^{\ell mn}_{x,x^2-y^2}$ &\l=0  \\
$t_{z,x^2-y^2}$   & \rule{0mm}{5mm}\n(\l$^2$-\m$^2$)[$\sqrt{3}/2$\ $t_{pd\sigma}$ - $t_{pd\pi}$] & $\beta^{\ell mn}_{z,x^2-y^2}$ &\n=0 
ot \l =\m\\
$t_{x,3z^2-r^2}$   & \rule{0mm}{5mm}\l\ $\{$[\n$^2$-(\l$^2$+\m$^2$)/2]\ $t_{pd\sigma}$-$\sqrt{3}$\ \n$^2$\ $t_{pd\pi}\}$ & 
$\beta^{\ell mn}_{x,3z^2-r^2}$ &\l=0  \\ \mr
$t_{xy,xy}$   & \rule{0mm}{5mm}3\l$^2$\m$^2$\ $t_{dd\sigma}$+(\l$^2$+\m$^2$-4\l$^2$\m$^2$)\ $t_{dd\pi}\cdots$ 
               & $\beta^{\ell mn}_{xy,xy}$ & -- \\
              & $\cdots$ +(\n$^2$+\l$^2$\m$^2$)\ $t_{dd\delta}$ &  &  \\
$t_{xy,yz}$   & \rule{0mm}{5mm}\l\n$\{$ 3\m$^2$\ $t_{dd\sigma}$+(1-4\m$^2$)\ $t_{dd\pi}\}$ + 
               (\m$^2$-1)\ $t_{dd\delta}$ & $\beta^{\ell mn}_{xy,yz}$ & \l=0 or \n=0 \\
$t_{xy,x^2-y^2}$   & \rule{0mm}{5mm}\l\m\ (\l$^2$-\m$^2$)\ $\{$ (3/2)$t_{dd\sigma}$-2\ $t_{dd\pi}$ + 
               (1/2)\ $t_{dd\delta}\}$ & $\beta^{\ell mn}_{xy,x^2-y^2}$ & \l=0 or \m=0 $\cdots$ \\
& & & $\cdots$ or \l=\m \\
$t_{yz,x^2-y^2}$   & \rule{0mm}{5mm}\m\n $\{$(3/2)(\l$^2$-\m$^2$)\ $t_{dd\sigma}$-[1+2(\l$^2$-\m$^2$)]\ $t_{dd\pi}$ $\cdots$ 
                & $\beta^{\ell mn}_{yz,x^2-y^2}$ & \m=0 or \n=0  \\
 & $\cdots$+ [1+1/2(\l$^2$-\m$^2$)]\ $t_{dd\delta}\}$ &  &   \\
$t_{xy,3z^2-r^2}$   & \rule{0mm}{5mm}$\sqrt{3}$\ \l\m\ $\{$ [\n$^2$-1/2(\l$^2$+\m$^2$)]\ $t_{dd\sigma}$-2\ \n$^2$\ $t_{dd\pi}$  
      $\cdots$          & $\beta^{\ell mn}_{xy,3z^2-r^2}$ & \l=0 or \m=0  \\
             & $\cdots$ +  1/2(1+\n$^2$)\ $t_{dd\delta}\}$ &  &   \\
$t_{x^2-y^2,x^2-y^2}$   & \rule{0mm}{5mm} 3/4(\l$^2$-\m$^2$)$^2$\ $t_{dd\sigma}$ +[\l$^2$+\m$^2$-(\l$^2$-\m$^2$)$^2$]\
$t_{dd\pi}$$\cdots$  & $\beta^{\ell mn}_{x^2-y^2,x^2-y^2}$ & --  \\
& $\cdots$ +[\n$^2$+1/4(\l$^2$-\m$^2$)$^2$]\ $t_{dd\delta}$ &  &   \\
$t_{x^2-y^2,3z^2-r^2}$   & \rule{0mm}{5mm}$\sqrt{3}$\ (\l$^2$-\m$^2$)$\{$ 
[\n$^2$/2-1/4(\l$^2$+\m$^2$)]\  $t_{dd\sigma}$ -\n$^2$\ $t_{dd\pi}$$\cdots$  & $\beta^{\ell mn}_{x^2-y^2,3z^2-r^2}$ &  \l=\m \\
& $\cdots$ +1/4(1+\n$^2$)\ $t_{dd\delta}$ &  &   \\
$t_{3z^2-r^2,3z^2-r^2}$   & \rule{0mm}{5mm} [\n$^2$-1/2(\l$^2$+\m$^2$)]$^2$\ $t_{dd\sigma}$ +3\n$^2$(\l$^2$+\m$^2$)\
$t_{dd\pi}$$\cdots$  & $\beta^{\ell mn}_{3z^2-r^2,3z^2-r^2}$ & --  \\
& $\cdots$ +3/4(\l$^2$+\m$^2$)$^2$\ $t_{dd\delta}$ &  &   \\ \br
\end{tabular}
\caption{Table showing the Slater-Koster parameters and the consequent symmetry factors $\beta^{\ell mn}_{LL'}$. The
parameters for which the zero condition have been omitted can be obtained from the last column by permuting the direction cosines.}
\label{tab1}
\end{table}

The above is an illustration, detailed arguments for the orbital symmetry has been discussed by Harrison \cite{harri}.

The \tref{tab1} provides the conditions for obtaining the orbital based symmetry
factor $\beta_{LL'}(R-R')$ for a lattice with cubic symmetry. We shall describe the
direction of $\vec{R}-\vec{R'}$ by the direction cosines ($\ell mn)$.  The table gives the
details for the calculation of the symmetry factor $\beta_{LL'}(\ell mn)$ on a cubic lattice.
A look at the figures \ref{fig2} and \ref{fig3} shows the reason for introduction of this factor.

For the $s$-orbital overlaps  the symmetry factor is always 1 (figure 1(a)). All six neighbours in the cubic lattice
in the directions (100),($\bar{\rm 1}$00),(010),(0$\bar{\rm 1}$0), (001) and (00$\bar{\rm 1}$) are equivalent with
the same overlap integrals $t_{ss\sigma}$ (only four are shown in the $xy$ plane). We may then reduce the lattice and retain only
the neighbour in the (100) direction and scale the overlap $t_{ss\sigma}$ by the appropriate weight (i.e. 2 in this case)  as discussed
earlier. For the $sp_x$,  however, the overlaps in the (010), (0$\bar{\rm 1}$0), (001) and (00$\bar{\rm 1}$) are
zero. This can be deduced immediately by noticing that the overlap products in these directions are positive for 
$x>0$ and negative for $x<0$. On reduction, although they are related to one
another by the symmetry operations of the cubic lattice,  we cannot club the neighbours
 in the (100) and ($\bar{\mathrm 1}$00) direction with those in the
(010),(0$\bar{\rm 1}$0),(001) and (00 $\bar{\rm 1}$) and will have to put these overlaps to be zero  by introducing
the symmetry factor $\beta_{LL'}(\ell mn)$.

\begin{equation}
\langle RL|H|R'L'\rangle \ \rightarrow \ \sqrt{\frac{{\cal W}_{R'}}{{\cal W}_R}} \ \beta_{LL'}(\ell mn) \ 
\langle RL|H|R'L'\rangle
\end{equation}

where $\beta_{LL'}(\ell mn)$  is 0 if $R'$ is 
a symmetric position of $R$ with respect to $L$ and $L'$, otherwise it is 1.

 \Fref{fig3} illustrates some of the overlaps involving the $d$ states
with $e_g$ and $t_{2g}$ symmetries.  With this simple reduction procedure the real space part can be reduced to
1/8-th, keeping only the sites in the ($x>0,y>0,z>0$) octant and suitably renormalizing the Hamiltonian matrix
elements as described earlier.

\begin{figure}
\centering
\epsfxsize=5 in \epsfysize=4in \rotatebox{0}{\epsfbox{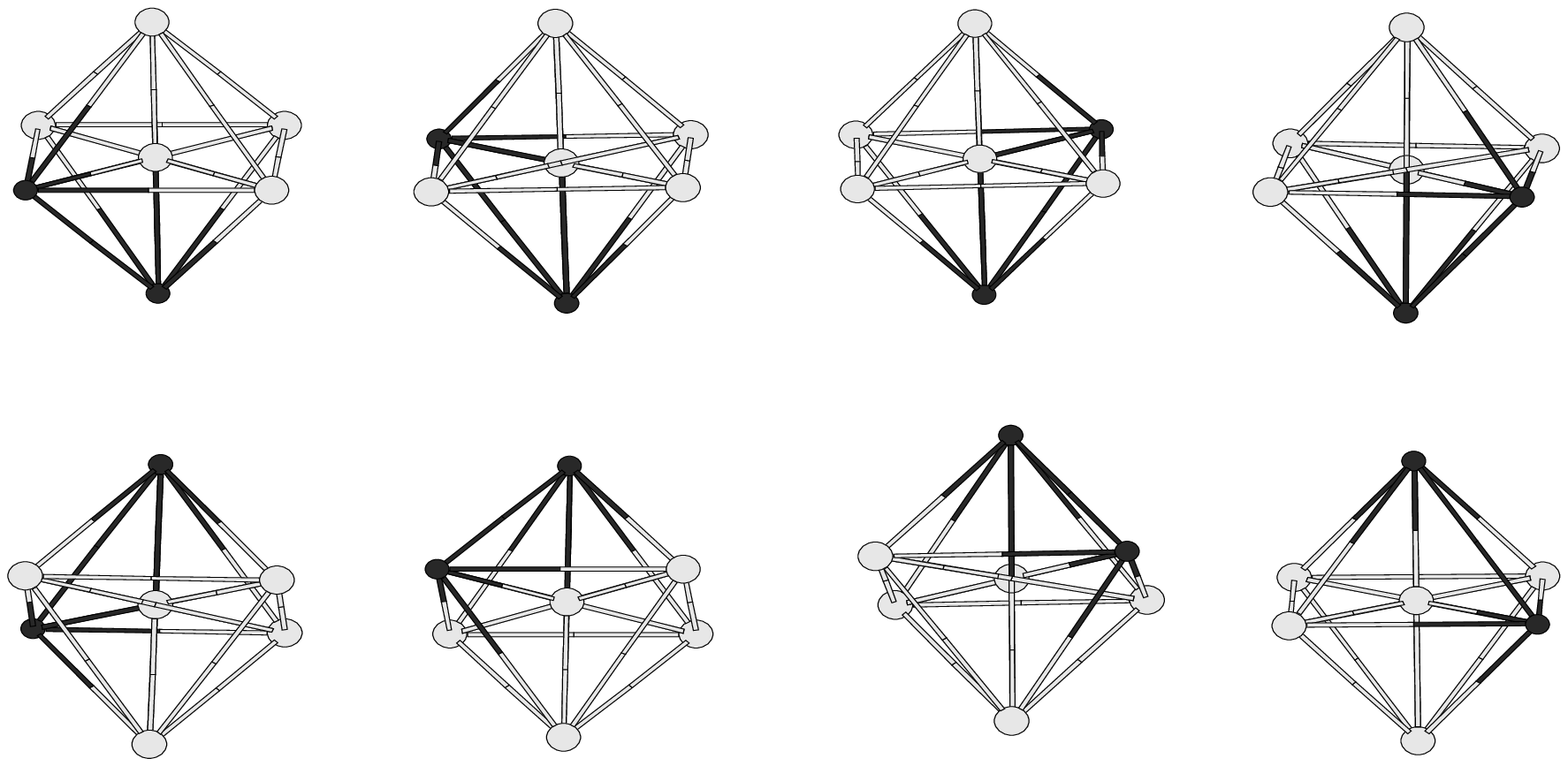}}
\vskip -1cm
\epsfxsize=5 in \epsfysize=2in \rotatebox{0}{\epsfbox{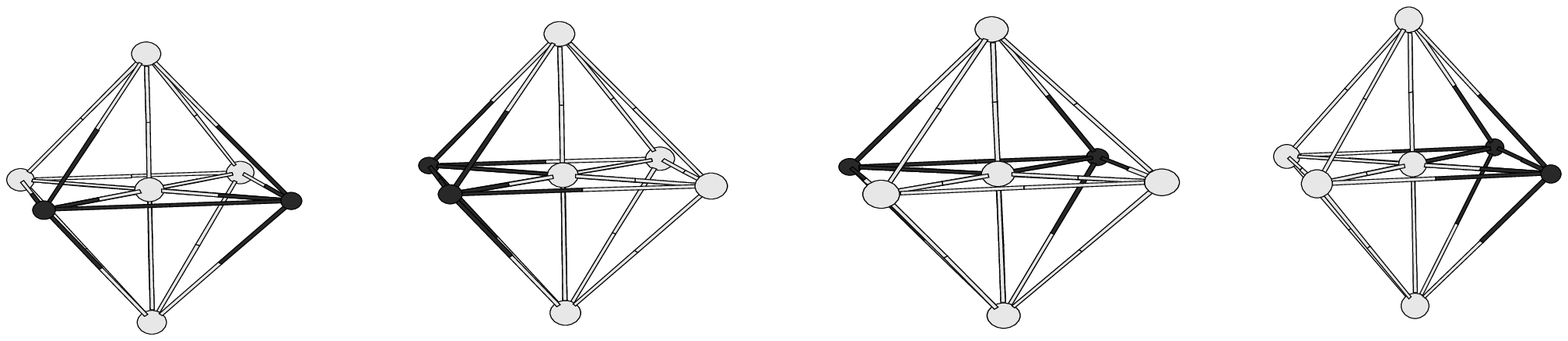}}
\vskip -1cm
\epsfxsize=4 in \epsfysize=2in \rotatebox{0}{\epsfbox{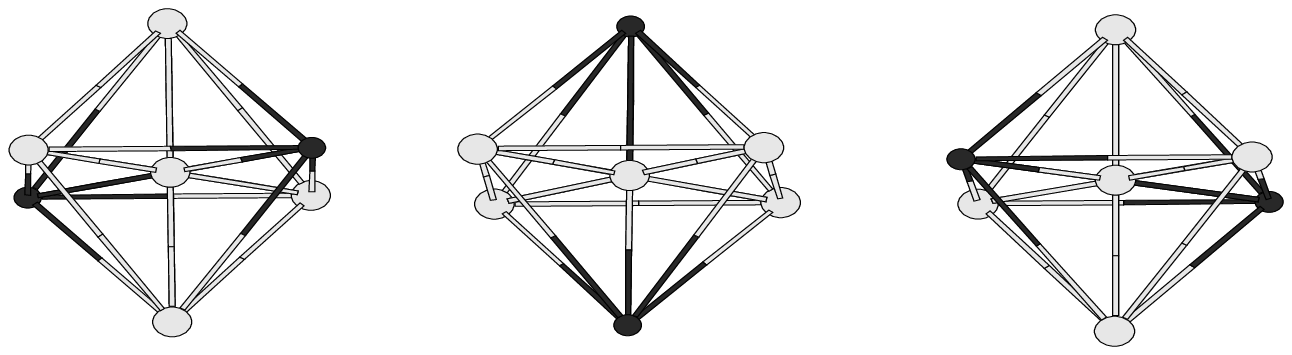}}
\raisebox{1cm}{\epsfxsize=1.5 in \epsfysize=1.5in \rotatebox{0}{\epsfbox{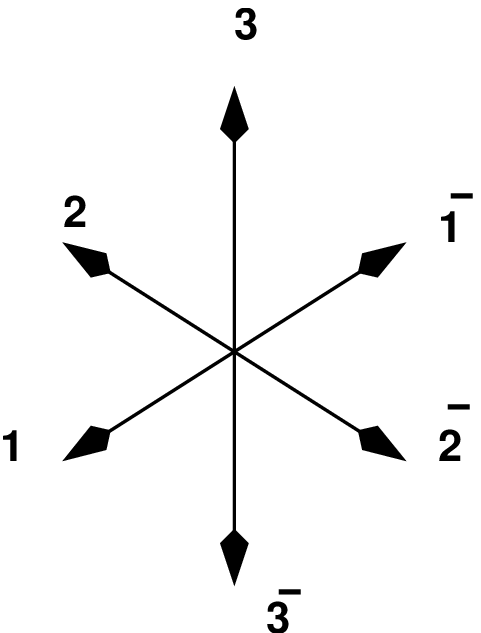}}}
\caption{Equivalent configurations on a cubic lattice. Configurations
labelled by $\uparrow$ are shown as light spheres, while those labelled
by $\downarrow$ are shown as dark spheres. The top twelve configurations
are equivalent, while the bottom three are equivalent. The lower right inset
shows the numbering scheme used for the lattice sites in the main text.}
\label{fig4}
\end{figure}

We still have not exhausted all the symmetries in the full augmented space. 
As discussed earlier, this space is a direct product of the real (lattice-orbital) space and
the configuration space which are disjoint. As a consequence the symmetry operations
apply independently to each of them. Since the disorder is homogeneous, the cardinality
sequence in configuration space itself has the symmetry of the underlying lattice.
To see this, let us look at the \fref{fig4} where we show a part of a cubic lattice
where the central site is occupied by a configuration labelled by $\uparrow$, while
two of the six nearest neighbours are occupied by configurations labelled by $\downarrow$,
and four of them by $\uparrow$s. We note that the twelve configurations in the first
three rows of the figure, where the two $\downarrow$s sit at distances $\sqrt{2}$ times
the lattice constant, are related to one another by the symmetry operations of the
cubic lattice. For example, the second to the fourth configurations
 on the top row can be obtained
from the first by the rotations ${\cal R}(\pi/2,\hat{z})$, ${\cal R}(\pi,\hat{z})$ 
and ${\cal R}(3\pi/2,\hat{z})$ respectively.  The configurations are described by {\sl cardinality
sequences} (as described earlier). The cardinality sequences for the four configurations 
on the top row of \fref{fig4} are : $\{1\bar{\rm 3}\}$, $\{2\bar{\rm 3}\}$,
 $\{\bar{\rm 1}\bar{\rm 3}\}$ and  $\{\bar{\rm 2}\bar{\rm 3}\}$. From the figure it is
easy to see that :

\[\fl \{2\bar{\rm 3}\} = {\cal R}(\pi/2,\hat{z})\  \{1\bar{\rm 3}\}\ ;\ 
 \{\bar{\rm 1}\bar{\rm 3}\} = {\cal R}(\pi,\hat{z})\  \{1\bar{\rm 3}\}\ \mbox{and}\ 
\{\bar{\rm 2}\bar{\rm 3}\} = {\cal R}(3\pi/2,\hat{z})\ \{1\bar{\rm 3}\}\]

\begin{figure}
\centering
\epsfxsize=3.5in\epsfysize=2.5in \raisebox{-0.1in}{\epsfbox{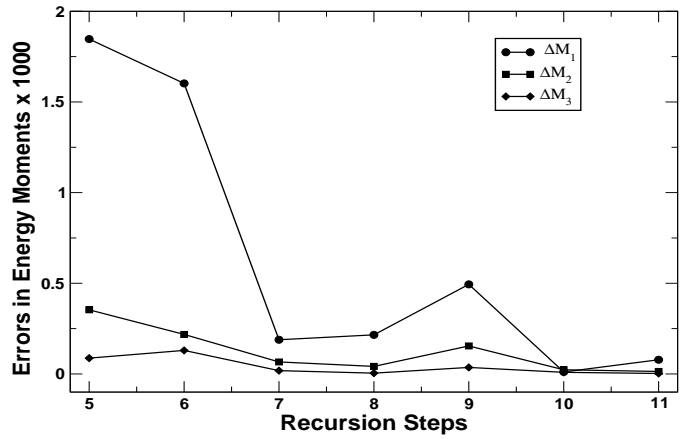}}
\caption{Convergence of the errors in energy moments of $Ag_{50}Pd_{50}$ as functions of the
recursion step.}
\label{fig5}
\end{figure}

This equivalence of the configurations on the lattice is quite independent of the symmetries of the Hamiltonian in real space discussed earlier.
 Thus, in augmented space, equivalent states are $|R \otimes 
\{\cal C\}\rangle$ and the set $|\Re R \otimes \Re \{\cal C\}\rangle$ for all different
symmetry operators $\Re$ of the underlying lattice. Again the symmetry of the orbitals
also rules out the operation of the Hamiltonian at certain symmetric positions discussed
earlier. 

\begin{figure}
\centering
\epsfxsize=4.0in \epsfysize=2.5in \rotatebox{0}{\epsfbox{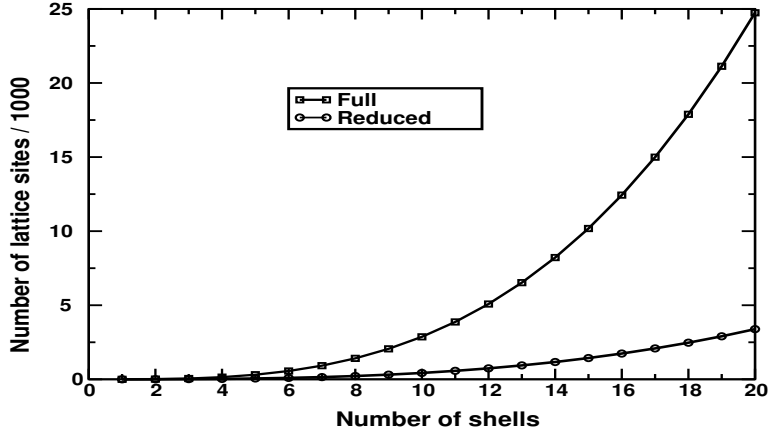}}
\caption{Showing how the number of lattice sites increase with increasing the number of shells in
the real space map and its reduced version. This is for an ordered fcc lattice.}
\label{fig6}
\end{figure}
 
\begin{table}[b]
\begin{center}
\begin{tabular}{|c|c|c|c|c|c|c|}
\hline
\multicolumn{3}{|c|}{} & \multicolumn{2}{|c|}{Full lattice}  & \multicolumn{2}{|c|}{Reduced lattice } \\
\hline
System & Shells & Steps & Sites & CPU time &  Sites & CPU time    \\
\hline
Model   & 20 & 20 & 24739 & 2.36  & 3385 & 0.54  \\
\hline
Ag & 20 & 30 & 24739 & 47.15 &  3385 & 15.83  \\
\hline
\end{tabular}
\label{tab2}
\caption{Comparison between system size and CPU time (in seconds) taken for recursion 
on a P4/256 machine for a full fcc lattice and the reduced lattice in real space.}
\end{center}
\end{table}

Once we have defined the reduced Hamiltonian, recursion on the reduced augmented space
with starting state $|u_1\rangle = |RL\otimes \{\emptyset\}\rangle$ gives the configuration
averaged Green function directly. 

In order to give the readers a flavour of the reduction in storage and CPU time, we have
carried out two sets of recursion calculations : First, a standard recursion on   ordered Ag 
on a fcc lattice, and second, an augmented space recursion with 15 seed points 
on a fcc alloy $Ag_{50}Pd_{50}$ both with and without
symmetry reduction. The calculations were done on   a P4 machine with 1.13 GHz clock speed and 256 MB RAM.

The first point to note is that convergence of the recursion technique is measured
by the convergence of the energy moments of the corresponding density of states
obtained from recursion \cite{vol35,atis}. These energy moments are defined by :

\[ M_n = \int_{-\infty}^{E_F} dE\ E^n n(E)\quad \mbox{where}\quad \int_{-\infty}^{E_F}
dE\ n(E) = n_e\]

\noindent where $n_e$ is the number of valence electrons. The convergence of these moments
 with the number of recursion steps $N$, from which the density of states $n(E)$ and
the Fermi energy $E_F$ is calculated, is reflected in the
{\sl errors} 

\[\Delta M_n^N = \vert M_n^N - M_n^{N-1}\vert \]
 
The \fref{fig5} shows the convergence of the first three moments for a $Ag_{50}Pd_{50}$
disordered alloy on a fcc lattice in a $spd$ TB-LMTO minimal basis set. It is clear from the figure
that for a convergence  within a specified error window one has to carry out recursion for a 
specific large number of steps determined by the error window. Symmetry reduction allows us to 
carry out recursion eactly over much larger number of steps than is possible for ordinary
recursion, given our computational resources. There was a
similar observation made in an earlier estimate of the errors of the recursion method \cite{atis}.
In order to carry out this many recursion steps we have to generate at least this many
shells exactly around the starting site in order to avoid surface-like effects.

\begin{figure}[t]
\centering
\epsfxsize=4.0in \epsfysize=2.5in \rotatebox{0}{\epsfbox{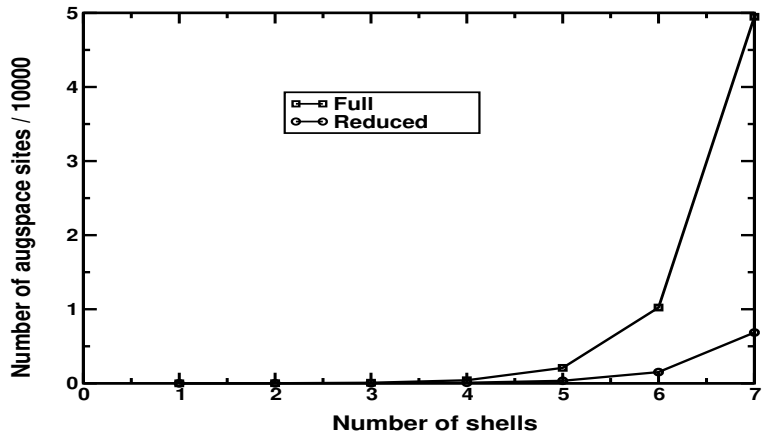}}
\caption{Showing how the number of lattice sites increase with increasing the number of shells in
an augmented space map on a disordered binary alloy on a  fcc lattice.}
\label{fig7}
\end{figure}

\begin{table}[b]
\begin{center}
\begin{tabular}{|c|c|c|c|c|c|c|}
\hline
\multicolumn{3}{|c|}{} & \multicolumn{2}{|c|}{Full lattice}  & \multicolumn{2}{|c|}{Reduced lattice } \\
\hline
 System & Shells & Steps & Sites & CPU time  & Sites & CPU time    \\
\hline
Model (50-50) & 7 & 7 & 49476 & 6.67 &  6856 & 1.21 \\
\hline
Ag$_{50}$Pd$_{50}$ & 7 & 11 & 49476 & 1136.6 &  6856 & 336.27 \\
\hline
\end{tabular}
\caption{Comparison between system size and time (in seconds) taken for recursion 
(using 15 seed points for Ag$_{50}$Pd$_{50}$) on a P4/256 machine for a full fcc 
lattice and the reduced lattice in augmented space.}
\end{center}
\label{tab3}
\end{table}

For the case of the ordered fcc lattice, 
the \fref{fig6} shows how the size of the map increases as we increase the number of nearest neighbour
shells from a starting site, both with and without reduction. Table 2 shows the details of the CPU time 
and storage space reduction for recursion after applying the symmetry reduction. We have carried out calculations both on a simple $s$-state system on a fcc lattice , as well as for
$Ag$ (with $spd$ minimal TB-LMTO basis) also on a fcc lattice. 

\begin{figure}
\centering
\epsfxsize=4in\epsfysize=2.5in\epsfbox{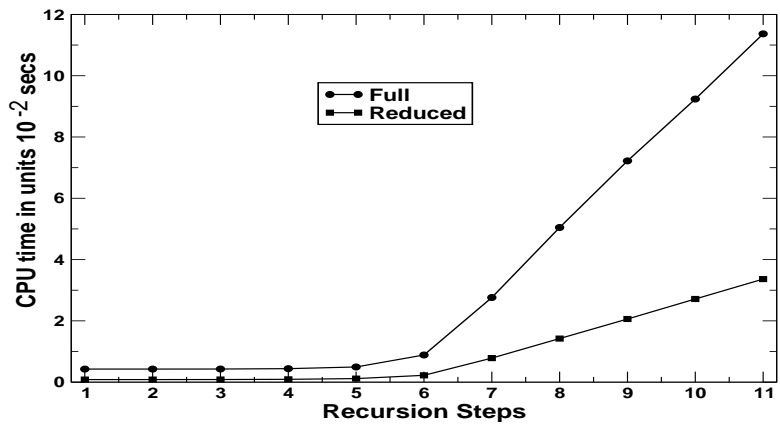}
\caption{CPU time taken per recursion step for a $Ag_{50}Pd_{50}$ disordered
binary alloy on a fcc lattice in a $spd$ TB-LMTO minimal basis set. The times
with and without symmetry reduction are shown.}
\label{time}
\end{figure}

For the calculation of a disordered binary alloy on a fcc lattice,
the \fref{fig7} shows the enormous decrease in  the size of the augmented space map after application
of symmetry reduction for a seven nearest neighbour map on a fcc lattice. \Tref{tab3} tabulates the
reduction in storage and CPU time for a 7 shell, 11 step recursion in augmented space carried over 15 seed points
using TB-LMTO potential parameters and structure matrix to built the Hamiltonian. The power of symmetry
reduction of storage space is more evident in this example. Such reduction will allow us to stretch our
nearest neighbour map up to 9-10 shells, stepping up our accuracy.  The \fref{time} indicates
the reduction in CPU time as we increase the number of recursion steps. Further, since the number of sites in
the map decrease, the number of individual applications of the $9\times9$ Hamiltonian also decrease
significantly, as do the number of operations involved in taking various inner products.
 This will reduce the inherent cumulative error of the recursion technique and lessen the probability of the
appearance of {\sl ghost bands} which often plague recursion calculations.

\section{Conclusions}
We have shown that recursion calculations can be carried out much faster and for
many more recursive steps exactly, if we perform the recursion on a subspace of
the original augmented space reduced by using the symmetries of both the underlying
lattice and random configurations on the lattice. This will allow us to obtain results
for disordered binary alloys with enhanced accuracy required for first-principles,
self-consistent, density functional based calculations. In this communication we have
described the details of the implementation of this symmetry reduction and
the modifications required in the standard recursion method. We propose using the symmetry
reduced version of the augmented space recursion in our future work on disordered alloys.

\end{document}